# The Impact of Thermal Conductivity and Diffusion Rates on Water Vapor Transport through Gas Diffusion Layers

S. Burlatsky[a]*, V. Atrazhev[b], M. Gummalla[a], D. Condit[a], and F. Liu[a]

[a] *United Technologies Research Center, East Hartford, CT 06108*

[b] *Russian Academy of Science, Institute of Biochemical Physics, Kosygin str. 4, Moscow, 119334, Russia*

\* Corresponding author: Phone: 1 860 610 7211, Fax: 1 860 610 7211

Email: BurlatSF@utrc.utc.com

**Abstract**

Water management in a hydrogen polymer electrolyte membrane (PEM) fuel cell is critical for performance. The impact of thermal conductivity and water vapor diffusion coefficients in a gas diffusion layer (GDL) has been studied by a mathematical model. The fraction of product water that is removed in the vapour phase through the GDL as a function of GDL properties and operating conditions has been calculated and discussed. Furthermore, the current model enables identification of conditions when condensation occurs in each GDL component and calculation of temperature gradient across the interface between different layers, providing insight into the overall mechanism of water transport in a given cell design. Water transport mode and condensation conditions in the GDL components depend on the combination of water vapor diffusion coefficients and thermal conductivities of the GDL components. Different types of GDL and water removal scenarios have been identified and related to experimentally-determined GDL properties.

*Keywords*: fuel cells, gas diffusion layer, thermal conductivity, water vapor, transport

# 1. Introduction

Optimal water management is important for polymer electrolyte membrane (PEM) fuel cell performance, durability and rapid start-up under frozen conditions. A low rate of water removal from the cathode reduces effective oxygen transport; while a high rate of water removal from the electrode may cause electrode dry-out resulting in low proton conductivity and low apparent exchange current, as well as a high membrane resistance.

The role of gas diffusion layer (GDL) is crucial to the overall water transport. Fundamental understanding of liquid water transport in the GDL and how to keep the bulk volume available for oxygen diffusion is necessary. Experimental determination of capillary pressure as a function of water saturation and mathematical formulation of Darcy's law have been the central issues in water management [1-6]. However, recent experimental and modeling findings indicate that the transport of liquid water in GDLs is a process of capillary fingering [7-11], which cannot be predicted by Darcy's law. Liquid water flows through GDL in the form of connected clusters, encountering dead ends due to the presence of varied diameter pores, and eventually percolates through pathway of least resistance [12]. Recently, Owejan et al. [13] found that vapor diffusion is the fundamental mechanism of water removal from the cathode catalyst layer through novel experimental techniques. It was shown that water flux of vapor driven by the thermal gradient in the cathode diffusion layer is shown to be sufficient to remove product water in vapor phase at high current densities in the gas delivery channels [13].

Thermal management and thermal property of the GDL are thus very important for high performance and durability [14-17]. Normally, in an operating fuel cell there is 3~5°C temperature gradient between the catalyst layer and gas channel [18, 19], which corresponds to 8~19% ΔRH at 75°C cell temperature. Furthermore, the heat released from water vapor condensation causes additional temperature difference within the electrode, creating a variation of water activity and, thus, ionic conductivity. The dissipation of reaction heat and water latent heat relies on GDL effective thermal conductivity, which strongly depends on the anisotropic packing and thermal contact of carbon particles and fibers. Interaction between mass and thermal transport makes favorable water management a challenge.

A promising water management approach is the use of porous bipolar plates, known as water-transport plates (WTPs) [20-24], developed by United Technologies. The WTPs perform two main functions. When there is excess water, the WTPs provide an evacuation path for liquid water to prevent flooding. When the gas streams are not saturated, the WTPs provide water to evaporate into the gas channels in order to humidify them [22]. The liquid water in the GDL is comprised of water streams flowing from the cathode to WTP and pendant or "stranded" water that is not connected to the water streams. In order to maximize oxygen transport, it is desirable to remove the major fraction of water in vapor phase to minimize the fraction of the GDL volume occupied by liquid water streams and to prevent vapor condensation and accumulation of pendant water in the GDL. The gradient of Water Vapor Partial Pressure (WVPP) is caused by the small temperature difference between the cathode and WTP resulting in the difference between the equilibrium WVPP in the WTP gas channel and that in the cathode gas pores. In this work, we use modelling, based on thermal conductivity measurement, to reveal a dynamics of product water vapor transport in a composite GDL comprised of substrates and micro-porous layer (MiPL) with different thermal conductivities and water vapor diffusion coefficients. This work also aims to provide a new physical model and clarify why an intermediate porous layer will aid water diffusive transport, as discussed by Owejan et al. [13].

## 2. Mathematical Formulations

### 2.1. Vapor Transport through cathode GDL

In this section, the governing equations of water vapor and thermal transport are introduced. The gas in the cathode-side GDL contains three components – oxygen, nitrogen and water vapor. The driving force for vapor transport through the GDL is the gradient of WVPP (see Appendix for details). Modeling results and experimental data indicate that in the present WTP design, the gas is fully saturated by water in the WTP near the inlet. The scenario of water transport (the fraction of water removed by vapor phase and condensation conditions in the GDL components) depends on the combination of water vapor diffusion coefficients, thermal conductivities of GDL components, and the boundary conditions at the GDL interfaces.

For simplification, we assume that no condensation or evaporation occurs within the GDL; therefore, there is no source or sink term for the thermal and mass transport equation. The water vapor in the GDL can thus be super-saturated, saturated, or under-saturated. Considering a 1D problem, the mass and thermal transport governing equations are described as follows

$$\frac{d}{dx}\left(D\frac{dC}{dx}\right) = 0 \tag{1}$$

$$\frac{d}{dx}\left(\lambda \frac{dT}{dx}\right) = 0 \tag{2}$$

where $D$ is vapor diffusion coefficient, and $\lambda$ is the thermal conductivity. For the boundary conditions, we assume that liquid water is in balance with the vapor at the cathode/MiPL interface (where x=1 in Figure 1) and the substrate/WTP interface (x=0). The water vapor concentration is higher at the cathode/MiPL interface than that in WTP because the temperature of the cathode electrode is higher than that of the WTP.

At the substrate and MiPL interface, we have the continuity of mass and heat fluxes:

$$D_S \frac{C_{SM} - C_0}{L_S} = D_M \frac{C_C - C_{SM}}{L_M} \tag{3}$$

$$\lambda_S \frac{T_{SM} - T_0}{L_S} = \lambda_M \frac{T_C - T_{SM}}{L_M} \tag{4}$$

therefore,

$$C_{SM} = C_C \frac{L_S}{L_S + (D_S/D_M)L_M} + C_0 \frac{(D_S/D_M)L_M}{L_S + (D_S/D_M)L_M} \tag{5}$$

$$T_{SM} = T_C \frac{L_S}{L_S + (\lambda_S/\lambda_M)L_M} + T_0 \frac{(\lambda_S/\lambda_M)L_M}{L_S + (\lambda_S/\lambda_M)L_M} \tag{6}$$

where $L$ is the thickness of each layer, and the subscripts 0, S, M, C, and SM indicate WTP, substrate, MiPL, cathode catalyst layer, and substrate/MiPL interface.

## 2.2. Type of GDL for Water Transport

In this section we shall derive the criteria for different types of GDLs in term of water transport mode. For common GDL materials, the estimated temperature difference between two sides is 2~3 °C at 1A/cm$^2$ (explained later). Within such a small temperature range, a linear relation between the saturated vapor concentration ($C^V$) and temperature can be reasonably assumed. At the two sides of the GDL, $C_C$ and $C_0$ are in equilibrium with liquid water, thus following this linear relation with temperature. If

$$\frac{D_S}{D_M} = \frac{\lambda_S}{\lambda_M} \tag{8}$$

Then, the calculated $C_{SM}$ and $T_{SM}$ as shown in Eq. 5 and Eq. 6 should also follow the same linear relation. This indicates that at the substrate/MiPL interface, $C_{SM}$ is equal to the saturated water vapor concentration at $T_{SM}$. On the other hand, if

$$\frac{D_S}{D_M} > \frac{\lambda_S}{\lambda_M} \text{ or } \frac{D_S}{\lambda_S} > \frac{D_M}{\lambda_M} \tag{9}$$

Then $C_{SM}$ is smaller while $T_{SM}$ is larger than those values obtained in the previous case. Thus in this case the water vapor is under saturated at the interface between the MiPL and substrate. To separate the impact of the boundary conditions from the impact of the GDL properties we define two types of GDL. The GDL belongs to Type 1 if Eq. 9 is fulfilled. Otherwise, the GDL belongs to Type 2. As shown in Figure 1 (a), the GDL belongs to Type 1 if the solution of the vapor diffusion equation predicts under-saturated vapor in the GDL. Otherwise, the GDL belongs to Type 2, as shown in Figure 1(b). The calculated WVPP distribution presented in Figure 1(b) is meta-stable. In the actual GDL with such WVPP distribution, water condensation would increase the saturation level and equilibrate the water vapor with the liquid water in the GDL bulk resulting in reduction of the water vapor transport rate and decrease in the limiting current.

In the above formulation, we assume a saturated water vapor concentration at the MiPL/electrode interface. In reality if the water vapor is under-saturated at the MiPL/electrode interface, it is much easier to achieve under-saturation in the bulk of the GDL. Eq. 9 is thus a very conservative criterion. All the GDLs within Type 1 have under-saturation water vapor in GDL bulk. Type 2 GDLs would have saturated, super-saturated and under-saturated vapor in the GDL bulk.

*2.3. Water Transport Scenarios*

Below we analyze water removal mode through the GDL and classify them into different scenarios as shown in Table 1. In Scenario 1, 100% of water generated in the cathode can be removed through the GDL in the gas phase and according to Scenario 2, less than 100% of water generated in the cathode can be removed in the gas phase. To calculate the fraction of water that the GDL removes through the gas phase, we compare the water generation rate to the maximum vapor flux through the GDL given that the water vapor is in equilibrium with liquid water at the GDL/WTP interface and that the temperature difference between the WTP and the cathode is fixed.

From Eq. 3, the vapor flux is

$$q_V = \frac{C_C - C_0}{\frac{L_S}{D_S} + \frac{L_M}{D_M}} = \frac{1}{R}\frac{P_C/T_C - P_0/T_0}{\frac{L_S}{D_S} + \frac{L_M}{D_M}} \tag{10}$$

where $P$ is the water vapor partial pressure. Assuming linear approximation of saturated vapor pressure in a small temperature range between $T_0$ and $T_C$, vapor pressure at the cathode is

$$P_C = P_0 + (T_C - T_0)\frac{\partial P_0}{\partial T} \tag{11}$$

Then Eq. 10 can be rewritten as

$$q_V = \frac{(T_C - T_0)}{RT_C} \frac{\frac{\partial P_0}{\partial T} - \frac{P_0}{T_0}}{\frac{L_S}{D_S} + \frac{L_M}{D_M}} \qquad (12)$$

Similar to Eq. 10, the heat flux is

$$q_H = \frac{T_C - T_0}{\frac{\lambda_S}{D_S} + \frac{\lambda_M}{D_M}} \qquad (13)$$

Combining Eqs. 12 and 13 gives

$$q_V = \frac{q_H \left(\frac{\lambda_S}{D_S} + \frac{\lambda_M}{D_M}\right)}{RT_C} \frac{\frac{\partial P_0}{\partial T} - \frac{P_0}{T_0}}{\frac{L_S}{D_S} + \frac{L_M}{D_M}} \qquad (14)$$

The heat flux can also be calculated by the total thermal balance at specific current density, $i$, i.e.,

$$q_H(i) = \alpha \left(Q_R^0 - Q_{W \to V}^0 \cdot \xi\right) \cdot i \qquad (15)$$

The specific reaction heat $Q_R^0$ and the latent water heat $Q_{W \to V}^0$ are calculated as functions of the specific reaction entropy and overpotential in Eq. 16 and Eq. 17.

$$Q_R^0 = T \frac{\Delta S_a}{2F} + T \frac{\Delta S_c}{4F} + \eta \qquad (16)$$

$$Q_{W \to V}^0 = \frac{\Delta H \cdot 18}{2F} \qquad (17)$$

Here $\eta$ is the overpotential, $\alpha$ is the fraction of heat that is removed from the cathode side, (Our calculations show that typically, $\alpha \approx 0.25 - 0.5$), $\Delta H$ is the water latent heat. At 1A/cm$^2$, $Q_{W \to V}^0 = 0.21 W/A$ and $Q_R^0 = 0.77 W/A$. At WTP temperature 65$^o$C, the cathode temperature $T_C$ and GDL temperature profile can thus be calculated using the thermal conductivity data shown in Table 3 and 4.

The fraction of water removed in the vapor phase, $\xi$, equals the ratio of the vapor flux to the total water generation rate. The maximum flux that can be removed from the cathode through the vapor phase in a Type 1 GDL is higher than the water production rate if

$$\xi = \frac{q_V}{i/2F} = \alpha\left(Q_R^0 - Q_{W \to V}^0 \cdot \xi\right)\frac{D}{\lambda}\frac{2F}{RT_C}\left(\frac{\partial P_0}{\partial T} - \frac{P_0}{T_0}\right) > 1 \tag{18}$$

where $\frac{L_S + L_M}{D} = \frac{L_S}{D_S} + \frac{L_M}{D_M}$, $\frac{L_S + L_M}{\lambda} = \frac{L_M}{\lambda_M} + \frac{L_S}{\lambda_S}$. $D$ and $\lambda$ are the effective vapor diffusion coefficient and thermal conductivity.

For Type 1 GDL we predict Scenario 1(a) if criterion (18) is valid and Scenario 2(a) otherwise. In the liquid-water-free GDL, $\alpha \approx 0.5$. Therefore Eq. 18 becomes

$$\xi = \frac{\beta}{1 + \beta\left(Q_{W \to V}^0 / Q_R^0\right)} > 1 \tag{19}$$

where

$$\beta = \frac{D}{\lambda} \cdot \frac{FQ_R^0}{RT_C}\left[\frac{\partial P_0}{\partial T} - \frac{P_0}{T_0}\right] \tag{20}$$

For Type 2 GDL, the vapor can condense in GDL. If criteria (18) is valid for the substrate ($D = D_S$, $\lambda = \lambda_S$), it is also valid for the MiPL ($D = D_M$, $\lambda = \lambda_M$), because for a Type 2 GDL the criteria (9) is not valid. As indicated in Figure 3a, the actual WVPP is lower than the saturated WVPP in the GDL. This case corresponds to Scenario 1(a). If the criteria (18) is not valid for the substrate, but it is valid for the MiPL, 100% of generated water is removed in the vapor phase through the MiPL, but some fraction condenses in the substrate. This fraction should be removed through the liquid phase, corresponding to Scenario 1(b) as shown in Figure 3b. If criterion (18) is not valid for the MiPL, less than 100% of water generated in the cathode is removed in the gas phase. The vapor in the MiPL is saturated, and vapor condenses in the GDL. This corresponds to the Scenario 2(b) in Figure 3c. The fraction of water removed though the vapor phase from the cathode in Scenario 2(b) is

$$\xi = \frac{\beta_1}{1 + \beta_1 \left( Q_{W \to V}^0 / Q_R^0 \right)}, \qquad (21)$$

where

$$\beta_1 = \frac{D_B}{\lambda_B} \cdot \frac{2\alpha F Q_R^0}{RT_C} \left[ \frac{\partial P_0}{\partial T} - \frac{P_0}{T_0} \right] \qquad (22)$$

## 3. Results and Discussion

Several sets of experimental measurements and estimates for the GDL thermal conductivity are currently available in the literature [18, 25]. Tables 3 and 4 show the thermal conductivities of substrate, MiPL and different composite combinations, determined by Laser Flash method [26] and a modified substrate hot plate method (MTM187) [27]. In the latter method for substrate conductivity, about 20 samples of substrate are stacked under 690 kPa (100 psi), and the heat flux through the stack is measured with a temperature gradient applied through the stack. The MiPL thermal conductivity was calculated by means of the inverse resistance law from the measured substrate and substrate/MiPL composite thermal conductivity. The measured thermal conductivities of wet-proofed and untreated Toray substrate, Toray/MiPL composites and different MiPLs surprisingly show that the measured MiPL thermal conductivity was about 30% higher than the one calculated by means of the inverse resistance law from the measured substrate and substrate/MiPL composite thermal conductivity. This may be due to the fact that the MiPL penetrates into the substrate. The thermal conductivity results in [27] are about equal for the untreated Toray substrate and 3x and 2x larger for the 10%TFE and 50%TFE MiPLs, respectively [26]. This may be attributed to lower thermal contact resistance between the stacked samples.

An analysis of the experimental results listed in Tables 3 and 4 implies Type 2 GDL and Scenario 1(b) (see Table 5) prevailed at 1A/cm$^2$. That means that the condensation is not expected in the MiPL and partial condensation is expected in the substrate as shown in Figure 3b. The MiPL can remove 100% of product water through the vapor phase at T=65 C. This is due to its lower thermal conductivity and therefore a locally high temperature zone is maintained near the cathode. For an untreated Toray substrate thickness of 0.2 mm, 10% TFE MiPL thickness of 50 μm and at a current density of 1 A/cm$^2$, the cathode temperature is higher than the WTP temperature by 2.6$^o$C using the data in Table 3 and by 2.2$^o$C using the data in Table 4.

The water transport scenario depends on the water vapor diffusion coefficient, which relies on porosity as shown in Table 6. The results for thermal conductivity obtained in Tables 3 and 4, imply that the fraction of water removed through the MiPL in the gas phase decreases from 100% to 50% with porosity decreasing from 0.5 to 0.25. If the MiPL porosity is 0.5 then 100% of water is removed by vapor phase for both 25μm and 75 μm MiPLs (Scenario 1(b)). The results for thermal conductivity obtained in Table 3, imply that 100% of water is removed through the MiPL in the gas phase even for the 10%TFE MiPL porosity of 0.25 and a MiPL thickness of 75 μm. However, the experimental results and theoretical models obtained for the current UTC WTP cell design indicate that the water generated in the cathode @1 A/cm$^2$ can be removed through vapor phase. In this case, the water vapor in the GDL can be under-saturated. Note that the theory developed in this paper may be valid for TWP only. Since the majority of reports for solid plate cells in the literature imply that there is a liquid/vapor equilibrium in the GDL, see [28, 29] and references therein.

A primary application for this work is to derive a preferred composite GDL that will maximize performance at high current density for WTP cell. To sustain enough mass transport, the major fraction of water should be removed through the gas phase and condensation in the GDL should be minimized. That means that the desired scenario is Scenario 2(a) and desired GDL type is Type 1 where liquid water and vapor are in equilibrium at the cathode-MiPL interface. In order to prevent the cathode from drying out and thereby sustaining high activity and low ionomer resistance, some fraction of generated water should be removed through the liquid phase as is the case for scenario 1(b) where some water is expected to condense in the substrate where the WTP can evacuate it (liquid water is wicked away and absorbed in the WTP more readily if the substrate is hydrophilic). For high current density with sufficient *ΔT*, all water for this scenario would be transported through the MiPL in the gas phase. At smaller current density with lower *ΔT*, a larger fraction of water could be removed through the liquid phase in both the MiPL and substrate. This, however, does not negatively impact on performance because the total water flux is lower than that at high current density. Thus, flooding in the MiPL would not be expected to occur.

## 4. Conclusions

A mathematical model has been developed to study water removal modes at the cathode in a porous-plate cell. The impact of thermal conductivity and diffusion coefficient on water transport in the GDL has been discussed. Assuming saturated vapor at both GDL interfaces it is concluded that a composite GDL structure (with MiPL) with low thermal conductivity in the vicinity of the cathode GDL interface helps to sustain under-saturated vapor in GDL bulk. Water transport (the fraction of water removed by vapor phase and condensation conditions in the GDL components) depends on the combination of water vapor diffusion coefficients and thermal conductivities of the GDL components. Different water removal scenarios were described (Table 1) and the criteria to predict the water removal scenario from GDL properties were provided (Table 2). The desirable fraction of water removed by vapor phase is about 90% for composite GDL, corresponding to Type 2 and Scenario 1(b) or 2(b), depending on the MiPL thickness and porosity (see Table 5). An analysis of thermal conductivity impact on water transport with 10%TFE and 50%TFE MiPLs predicts that more than 90% of product water (@ 1A/cm2) is transported through these MiPLs in the vapor phase. Condensation of liquid water is not expected anywhere in Type 1 composite GDL as opposed to Type 2 composite GDL where condensation in substrate and/or MiPL is dependent on the water removal scenario.

**Appendix A**:

*A 1: The Driving Force of the Vapor Transport in GDL*

At the GDL/WTP interface, water vapor is in balance with liquid water. The water vapor chemical potential at the GDL/WTP interface is equal to the liquid water chemical potential in the WTP. However, the chemical potential of the liquid water in the cathode is higher than the chemical potential of the liquid water in WTP because the temperature of the cathode is higher than the temperature of the WTP. This implies that the water vapor chemical potential in the cathode is higher than the water vapor chemical potential at the GDL/WT interface.

The temperature and chemical potential gradients cause the thermo-diffusion and the diffusion fluxes the GDL.

To understand the mechanism of the vapor transport in the liquid-water-free GDL we use a simple cartoon, depicted in Figure A1. The cathode and WTP are shown as two vessels, partially filled with liquid water, GDL is shown as a liquid-water-free pipe between these two vessels. Liquid water is in equilibrium with vapor in both vessels. The temperature of the left hand vessel, $T_1$, is higher than the temperature of the right hand vessel, $T_2$, $T_1 > T_2$.

At P = 1 atm, the water vapor can be treated as the ideal gas. The chemical potential of one mole of ideal gas is

$$\mu_v(P,T) = \mu_v^0(T) + RT \cdot \ln(P) \tag{A1}$$

where $P = p/p_0$ is the normalized pressure, $p_0$ is the partial pressure of gas under the standard condition, $\mu_v^0(T)$ is standard chemical potential of the gas. The concentration of a component of the ideal gas mixture is

$$C = \frac{P}{RT} \tag{A2}$$

Here $P$ is the partial pressure of a gas component. The chemical potential of the gas component is a function of its partial pressure $P$ (or concentration $C$) and of the gas temperature $T$ The chemical potential of liquid water is a function of the temperature only

$$\mu_w = \mu_w(T) \tag{A3}$$

In liquid water/water vapor equilibrium system, the water vapor chemical potential is equal to the liquid water chemical potential. From Equations (A1) and (A5), we obtain the following Equation for the sated vapor pressure in the vessels

$$\mu_w(T) = \mu_v^0(T) + RT\ln(P) \tag{A4}$$

Solving it with respect to *P*, we obtain the following dependence of the saturated vapor pressure on temperature

$$P(T) = \exp\left\{\frac{\mu_w(T) - \mu_v^0(T)}{RT}\right\} \tag{A5}$$

Equation (A5) indicates that the vapor chemical potential and pressure in both vessels depend only on vessel temperature. Using the Equation (A2) we obtain the following Equation for saturated vapor concentration

$$C(T) = \frac{\exp\left\{\dfrac{\mu_w(T) - \mu_v^0(T)}{RT}\right\}}{RT} \tag{A6}$$

There is no liquid water in the pipe => vapor concentration and temperature in a pipe should be treated as independent variables. Vapor concentration and temperature in the left hand side vessel is higher than that in the right hand side vessel. This results in the chemical potential gradient and in the diffusion flux as shown in following Section.

*A 2: Fick's Law*

If the concentration and temperature gradients in the gas are small, the diffusion flux, *j*, of one of the gas components, such as water vapor, is a linear function of the gradient of the chemical potential of this component and of the temperature gradient in the gas [28].

$$j = -\alpha \cdot \nabla\mu_i - \beta \cdot \nabla T \tag{A7}$$

Equation (A4) using (A2) can be re-written in the form

$$\mu_w(T) = \mu_v^0(T) + RT\ln(C_i RT) \tag{A8}$$

, where the local water vapor chemical potential $\mu_i$ is a function of the local water vapor concentration $C_i$ and of the local gas temperature $T$. The gradient of the chemical potential is a function of the concentration and temperature gradients

$$\nabla \mu = \left(\frac{\partial \mu}{\partial C_i}\right)_T \nabla C_i + \left(\frac{\partial \mu}{\partial T}\right)_C \nabla T \tag{A9}$$

Substituting (A9) into (A7), we obtain the following Equation for the flux

$$j = -\alpha \left(\frac{\partial \mu}{\partial C_i}\right)_T \nabla C_i - \left[\alpha \left(\frac{\partial \mu}{\partial T}\right)_C + \beta\right] \nabla T \tag{A10}$$

Using the standard definitions of the transport rate coefficients

$$D = \alpha \left(\frac{\partial \mu}{\partial C_i}\right)_T \tag{A11}$$

$$\frac{k_T D}{T} = \alpha \left(\frac{\partial \mu}{\partial T}\right)_C + \beta \tag{A12}$$

we finally obtain the following transport rate Equation

$$j = -D\left(\nabla C_i + \frac{k_T}{T} \nabla T\right) \tag{A13}$$

Here $k_T$ is a thermo-diffusion ratio. The diffusion coefficient $D$ depends on the concentrations of all component of the gas and on the temperature. However, if the concentration and temperature gradients are small (as in our case) it can be treated as a constant.

*A 3: Kinetic coefficients in the transport equation*

The diffusion and thermo-diffusion coefficient cannot be calculated using thermodynamic methods. Kinetic theory [29] provides the following Equations for these coefficients in the gas phase

$$D = \frac{kT}{3P}\left\langle \frac{v}{\sigma_t} \right\rangle \tag{A14}$$

$$k_T = C_i T \frac{\partial}{\partial T} \ln\left( \frac{\langle v/\sigma_t \rangle}{kT} \right) \tag{A15}$$

where $v$ is the thermal velocity of the gas component molecule, $\sigma_t$ is the transport collision cross-section of the gas component molecule ($\sigma_t$ practically does not depend on the velocity). The mean thermal velocity is a function of the temperature $\langle v \rangle = (kT/m_i)^{1/2}$, where $m_i$ is the molecular weight. Thus, for the thermo-diffusion ratio we obtain the following Equation

$$k_T = \frac{C_i}{2} \tag{A16}$$

Finally, the flux of the gas component is

$$j = DC_i\left[ \frac{\nabla C_i}{C_i} - \frac{\nabla T}{2T} \right] \tag{A17}$$

*A 4: Fick's flux versus thermo-diffusion in the GDL*

Here we compare the first (diffusion) and second (thermo-diffusion) terms in the right hand side of Equation (A17) using the typical GDL conditions in PEM. The first term in the right hand side of Equation (A17) is

$$\frac{\nabla C_v}{C_v} \approx \frac{C_C - C_0}{LC_0} \tag{A18}$$

Here $L$ is the GDL thickness. Using the equation of state of ideal gas, Equation (A2), we express the vapor concentration in terms of the partial pressure

$$\frac{\nabla C_v}{C_v} \approx \frac{P_C - P_0}{LP_0} \tag{A19}$$

Having in mind, that the water vapor is sated in the cathode and at the cathode/WTP interface, we obtain the following Equation for the pressure gradient

$$\frac{P_C - P_0}{LP_0} \approx 2\frac{\partial P_0}{\partial T}\frac{T}{P_0}\frac{\nabla T}{2T} \qquad (A20)$$

Taking advantage of Equation (A20), we obtain the following Equation for the ratio of the diffusion/thermo-diffusion fluxes

$$\frac{\nabla C_i}{C_i} \bigg/ \frac{\nabla T}{2T} = 2\frac{\partial P_0}{\partial T}\frac{T}{P} \qquad (A21)$$

At $65^0 C$,

$$\frac{\nabla C_i}{C_i} \bigg/ \frac{\nabla T}{2T} = \frac{2 \cdot 10^4 \cdot 340}{25 \cdot 10^4} = 27.2 \qquad (A22)$$

Equations (A21) and (A22) implies that, under typical PEM operational conditions, the second term (thermo-diffusion) in Equation (A17) is much smaller than the first one (diffusion) and that the conventional diffusion Equation

$$j = D\nabla C_i \qquad (A23)$$

is applicable to the water vapor transport in GDL.

**Nomenclature**

        **Parameters**

| | |
|---|---|
| $D$ | Water vapor diffusion coefficient in porous medium, $cm^2/s$ |
| $D_{eff}$ | Effective water vapor diffusion coefficient in substrate+MiPL composite, $cm^2/s$ |
| $F$ | Faraday's constant, $96487\,C/mol$ |
| $\Delta H$ | Water latent heat, $J/g$ |
| $i$ | Current density, $A/cm^2$ |
| $L$ | Thickness, $cm$ |
| $P_0$ | Saturated water vapor pressure at the WPT coolant temperature, $din/cm^2$ |
| $Q^0_{W \to V}$ | Latent water heat per unit current density, $W/A$ |
| $Q^0_R$ | Specific reaction heat per unit current density, $W/A$ |
| $R$ | Ideal gas constant, $8.7 \cdot 10^7\,erg/mol \cdot K$ |
| $T$ | Temperature, $K$ |
| $T_0$ | WTP coolant temperature, $K$ |
| $S$ | Specific reaction entropy, $J/mol \cdot K$ |
| $\alpha$ | Fraction of heat that is removed by the cathode-side WTP |
| $\eta$ | Overpotential, $V$ |
| $\lambda$ | Thermal conductivity, $W/cm \cdot K$ |
| $\lambda_{eff}$ | Effective thermal conductivity of substrate+MiPL composite, $W/cm \cdot K$ |
| $\xi$ | Fraction of water removed through the vapor phase |

**Subscripts**

| | |
|---|---|
| $A$ | Anode |
| $SM$ | Interface between the substrate and MiPL |
| $C$ | Cathode |
| $S$ | Substrate |
| $M$ | Membrane |

**List of Figures:**

**List of Tables:**

Table 1. Possible Scenarios of water removal through the GDL

| GDL Type | Scenario of water dynamics | Condensation in the GDL | Fraction of water removed through the gas phase | Liquid water streams in the GDL | Drops and puddles in the GDL |
|---|---|---|---|---|---|
| Type 1 | Scenario 1(a) | No condensation Under-saturated vapor in the GDL | 100% | NO | NO |
| | Scenario 2(a) | | Less than 100% | **YES** | NO |
| Type 2 | Scenario 1(a) | | 100% | NO | NO |
| | Scenario 1(b) | Condensation in the substrate Under-saturated vapor in the MiPL | Less than 100% | NO | YES in the substrate |
| | Scenario 2(b) | Condensation in the substrate Saturated vapor in the MiPL | Less than 100% | **YES** | YES in the substrate |

Table 2. Criteria determining the scenario of water removal through the substrate

| GDL Type | Type 1 | | Type 2 | | |
|---|---|---|---|---|---|
| Scenario | Scenario 1(a) | Scenario 2(a) | Scenario 1(a) | Scenario 1(b) | Scenario 2(b) |
| Criteria (9) | **True** | **True** | False | False | False |
| Criteria (18) for the substrate | N/A | N/A | **True** | False | False |
| Criteria (18) for the MiPL | N/A | N/A | **True** | **True** | False |
| Criteria (18) for the whole GDL | **True** | False | N/A | N/A | N/A |

Table 3. Laser Flash Thermal Conductivity measurement results [26]

| Sample | thickness $L$ @ 25°C (mm) | bulk density $\rho$ @ 25°C (g/cm$^3$) | temperature (°C) | diffusivity D (cm$^2$/s) | specific heat $C_p$ (J/g-K) | conductivity $\lambda$ (W/m-K) |
|---|---|---|---|---|---|---|
| **Wet-proofed (30%TFE) Toray only** | | | | | | |
| w/ penetration model | 0.19 | 0.62 | 65 | 0.0374 | 0.922 | 2.13 |
| w/ std. model | 0.19 | 0.62 | 65 | 0.0566 | 0.922 | 3.23 |
| **Nominal 3 mil, 50% TFE MiPL+ Wet-proofed Toray** | | | | | | |
| w/ penetration model | 0.25 | 0.61 | 65 | 0.00584 | 0.870 | 0.310 |
| w/ std. model | 0.25 | 0.61 | 65 | 0.00598 | 0.870 | 0.318 |
| **Nominal 1 mil, 50% TFE MiPL only** | | | | | | |
| w/ penetration model | 0.045 | 0.59 | 65 | 0.00090 | 1.07 | 0.057 |
| w/ std. model | 0.045 | 0.59 | 65 | 0.00097 | 1.07 | 0.062 |
| **Untreated Toray only** | | | | | | |
| w/ penetration model | 0.20 | 0.41 | 65 | 0.0343 | 0.756 | 1.06 |
| w/ std. model | 0.20 | 0.41 | 65 | 0.0545 | 0.756 | 1.69 |
| **Nominal 1 mil, 10% TFE MiPL+untreated Toray** | | | | | | |
| w/ penetration model | 0.25 | 0.50 | 65 | 0.00501 | 0.862 | 0.214 |
| w/ std. model | 0.25 | 0.50 | 65 | 0.00521 | 0.862 | 0.222 |
| **Nominal 1 mil, 10% TFE MiPL only** | | | | | | |
| w/ penetration model | 0.070 | 0.47 | 65 | 0.00083 | 0.887 | 0.035 |
| w/ std. model | 0.070 | 0.47 | 65 | 0.00096 | 0.887 | 0.040 |

Table 4. Laser Flash versus Modified Substrate method (MTM187) results [27]

| | **Laser Flash Thermal Conductivity (W/m/k)** | **Through-Plane Thermal Conductivity (W/m/k)** |
|---|---|---|
| **10% MiPL** | 0.035 | 0.097 |
| **50% MiPL** | 0.057 | 0.097 |
| **Toray Substrate** | 1.06 | 0.798 |

Table 5. Scenarios of vapor transport for different sets of GDL parameters (indexes: S – substrate, M – MiPL, $\alpha$ - heat fraction removed through cathode WTP).

| %TFE (MiPL thickness mil) | $\alpha$ | Parameters | Criteria (9) | Criteria (18) for the substrate | Criteria (18) for the MiPL | Scenario |
|---|---|---|---|---|---|---|
| **10%(1)** | 0.36 | $\lambda_S = 1.06*10^{-2}$ W/cm*K<br>$\lambda_M = 0.035*10^{-2}$ W/cm*K<br>$\varepsilon_M=0.78$, $L_M = 25\mu m$ | False<br>1.3>47 | False<br>0.2>1 | True<br>6.8>1 | 1(b) |
| **50%(1)** | 0.41 | $\lambda_S = 1.06*10^{-2}$ W/cm*K<br>$\lambda_M = 0.057*10^{-2}$ W/cm*K<br>$\varepsilon_M=0.72$, $L_M = 25\mu m$ | False<br>1.3>26 | False<br>0.22>1 | True<br>4.2>1 | 1(b) |
| **10%(3)** | 0.22 | $\lambda_S = 1.06*10^{-2}$ W/cm*K<br>$\lambda_M = 0.035*10^{-2}$ W/cm*K<br>$\varepsilon_M=0.78$, $L_M = 75\mu m$ | False<br>1.3>47 | False<br>0.11>1 | True<br>4.1>1 | 1(b) |
| **50%(3)** | 0.28 | $\lambda_S = 1.06*10^{-2}$ W/cm*K<br>$\lambda_M = 0.057*10^{-2}$ W/cm*K<br>$\varepsilon_M=0.72$, $L_M = 75\mu m$ | False<br>1.3>26 | False<br>0.14>1 | True<br>2.9>1 | 1(b) |
| **50%(1)** | 0.46 | $\lambda_S = 0.8*10^{-2}$ W/cm*K<br>$\lambda_M = 0.097*10^{-2}$ W/cm*K<br>$\varepsilon_M=0.72$, $L_M = 25\mu m$ | False<br>0.17>1.51 | False<br>0.32>1 | True<br>2.8>1 | 1(b) |
| **50%(3)** | 0.37 | $\lambda_S = 0.8*10^{-2}$ W/cm*K<br>$\lambda_M = 0.097*10^{-2}$ W/cm*K<br>$\varepsilon_M=0.72$, $L_M = 75\mu m$ | False<br>0.17>1.51 | False<br>0.26>1 | True<br>2.3>1 | 1(b) |

Table 6. Scenarios of vapor transport for different MiPL porosity (indexes: S – substrate, M – MiPL, $\alpha$ - heat fraction removed through cathode WTP).

| %TFE (MiPL thickness mil) | $\alpha$ | Parameters | Criteria (9) | Criteria (18) for the substrate | Criteria (18) for the MiPL | Scenario |
|---|---|---|---|---|---|---|
| **10%(1)** | 0.36 | $\lambda_S = 1.06*10^{-2}$ W/cm*K<br>$\lambda_M = 0.035*10^{-2}$ W/cm*K<br>$\varepsilon_M=0.25$, $L_M = 25\mu m$ | False<br>1.3>8.5 | False<br>0.2>1 | **True**<br>1.2>1 | 1(b) |
| **10%(1)** | 0.36 | $\lambda_S = 1.06*10^{-2}$ W/cm*K<br>$\lambda_M = 0.035*10^{-2}$ W/cm*K<br>$\varepsilon_M=0.5$, $L_M = 25\mu m$ | False<br>1.3>8.5 | False<br>0.2>1 | **True**<br>3.5>1 | 1(b) |
| **10%(3)** | 0.22 | $\lambda_S = 1.06*10^{-2}$ W/cm*K<br>$\lambda_M = 0.035*10^{-2}$ W/cm*K<br>$\varepsilon_M=0.25$, $L_M = 75\mu m$ | False<br>1.3>8.5 | False<br>0.11>1 | False<br>0.75>1 | 2(b) |
| **10%(3)** | 0.22 | $\lambda_S = 1.06*10^{-2}$ W/cm*K<br>$\lambda_M = 0.035*10^{-2}$ W/cm*K<br>$\varepsilon_M=0.5$, $L_M = 75\mu m$ | False<br>1.3>8.5 | False<br>0.11>1 | **True**<br>2.1>1 | 1(b) |
| **50%(1)** | 0.4 | $\lambda_S = 0.8*10^{-2}$ W/cm*K<br>$\lambda_M = 0.097*10^{-2}$ W/cm*K<br>$\varepsilon_M=0.25$, $L_M = 25\mu m$ | False<br>1.7>3. | False<br>0.3>1 | False<br>0.57>1 | 2(b) |
| **50%(3)** | 0.37 | $\lambda_S = 0.8*10^{-2}$ W/cm*K<br>$\lambda_M = 0.097*10^{-2}$ W/cm*K | False<br>1.7>3. | False<br>0.26>1 | False<br>0.46>1 | 2(b) |
| | | $\varepsilon_M=0.25$, $L_M = 75\mu m$ | | | | |
| **50%(1)** | 0.4 | $\lambda_S = 0.8*10^{-2}$ W/cm*K $\lambda_M = 0.097*10^{-2}$ W/cm*K $\varepsilon_M=0.5$, $L_M = 25\mu m$ | False 1.7>8. | False 0.32>1 | **True** 1.6>1 | 1(b) |
| **50%(3)** | 0.37 | $\lambda_S = 0.8*10^{-2}$ W/cm*K $\lambda_M = 0.097*10^{-2}$ W/cm*K $\varepsilon_M=0.5$, $L_M = 75\mu m$ | False 1.7>8. | False 0.26>1 | **True** 1.3>1 | 1(b) |

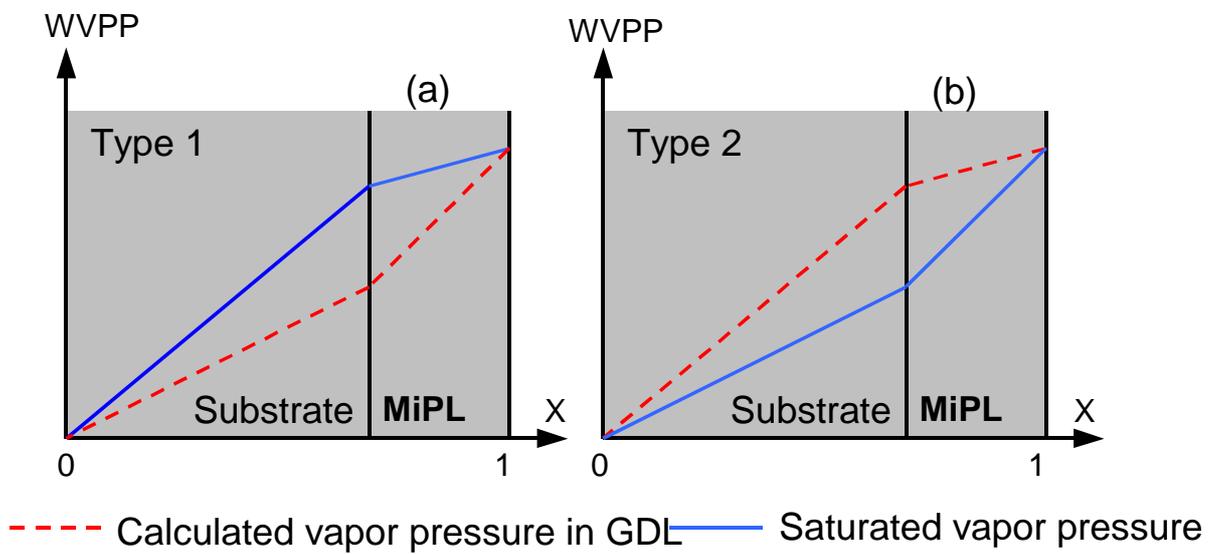

Figure 1 The WVPP distribution (dashed red line) calculated from the vapor diffusion equations is compared to the WVPP distribution of the saturated vapor (solid blue line). The GDL belongs to Type 1 if the red line is located below the blue line (a) and it belongs to Type 2 otherwise (b). The diffusion equations are solved under the condition that temperature of the cathode/GDL (x=1) interface is higher than that of the GDL/WTP interface (x=0) and the vapor is saturated at these boundaries.

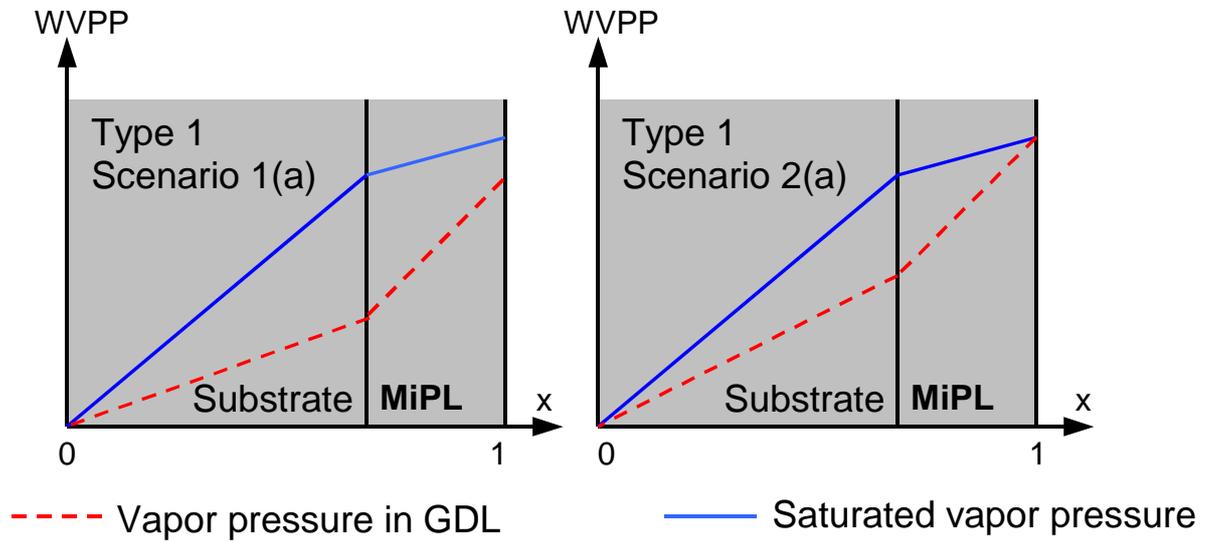

Figure 2 The WVPP distribution in the GDL of Type 1 is compared to the saturated WVPP distribution. The vapor in the GDL is under-saturated and there is no condensation in any GDL component. In Scenario 1(a), the vapor at the cathode/GDL interface is under-saturated and 100% of water generated in the cathode can be removed through the vapor phase. In Scenario 2(a), the vapor at the cathode/GDL interface is saturated and less than 100% of water generated in the cathode can be removed through the vapor phase.

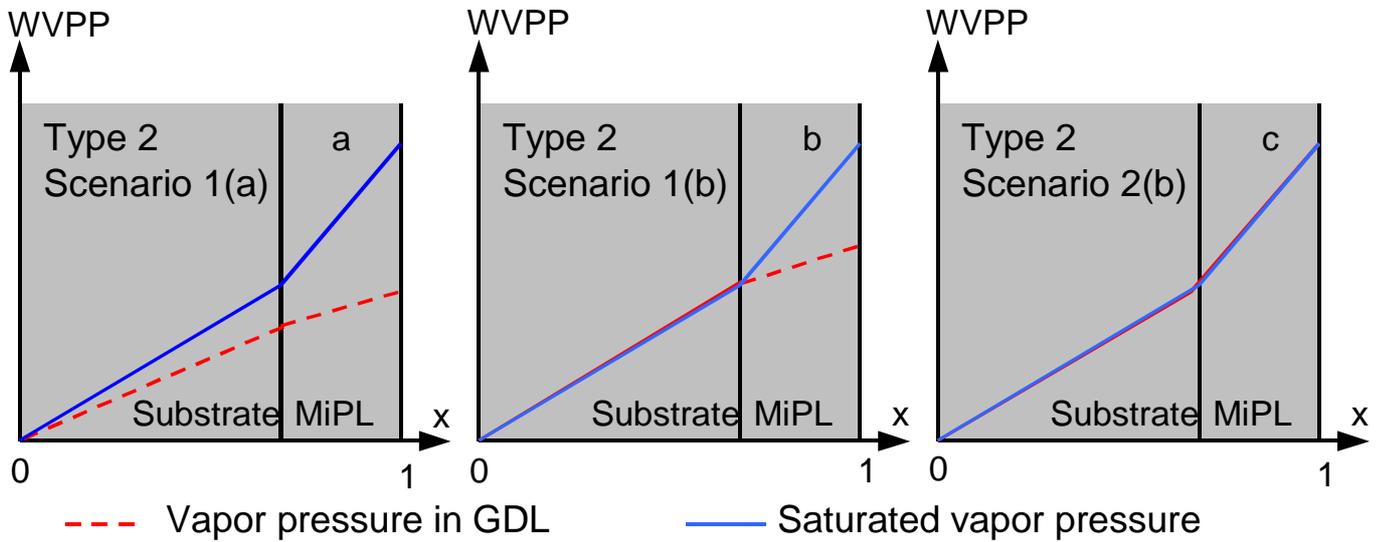

Figure 3 Actual WVPP distribution is compared to the saturation vapor pressure in the Type 2 GDL. In Scenario 1(a), the actual WVPP is lower than the saturated vapor pressure. In Scenario 1(b), the actual WVPP is equal to the saturated vapor pressure in the substrate. In Scenario 2(b) the actual WVPP is equal to the saturated vapor pressure in the substrate and MiPL.

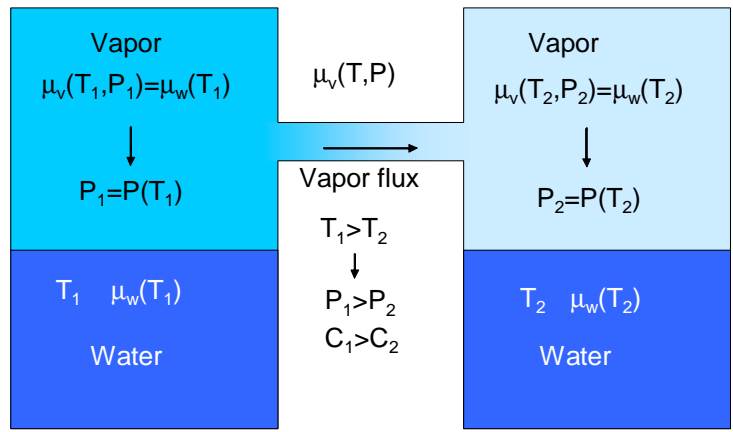

Figure A1: In both of the vessels, liquid water is in equilibrium with the vapour and the vapour partial pressure (concentration) is a function of the temperature of the vessel. There is no liquid water in a pipe connecting the vessels so the water vapor concentration and the temperature in the pipe are independent.